\pdfoutput=1 
\documentclass{JINST}

\def\ifmath#1{\relax\ifmmode #1\else $#1$\fi}%
\def\fbi{\mbox{fb$^{-1}$}}%

\title{Radiation background with the CMS RPCs at the LHC}

\author{Silvia Costantini\thanks{Corresponding author.} \\
on behalf of the CMS Collaboration\\

\llap{}University of Ghent, Department of Physics and Astronomy, \\
	Proeftuinstraat 86, B-9000, Ghent, Belgium\\

E-mail: \email{silvia.costantini@cern.ch}}

\abstract{The Resistive Plate Chambers (RPCs) are employed in the 
CMS experiment
at the LHC as dedicated trigger system both in the barrel and in the endcap.
This note presents results of the 
radiation background measurements performed with the 2011 and 2012
proton-proton collision data collected by CMS. Emphasis is given
to the measurements of the background distribution inside the RPCs.
The expected background rates during the future running of the LHC 
are estimated both from extrapolated measurements and from simulation. 
}


\begin{document}

\section{Introduction}

Three years (2010-2012) of proton-proton (pp) running at the 
Large Hadron Collider (LHC)
have provided the opportunity to extensively study the radiation background 
and therefore to obtain a very detailed knowledge of the  background
distribution in the Muon system of the Compact Muon Solenoid (CMS)~\cite{cms1, cms2} detector.
The measurements presented in this note have been performed with
2011 and 2012 
runs of  proton-proton collisions 
at a center-of-mass energy of 7 TeV, and at 8 TeV, respectively,
The maximal instantaneous luminosity reached
$7.7 \cdot 10^{33}$ cm$^{-2}$ s$^{-1}$ in 2012.
Previous results with CMS 2010 data at 7~TeV are reported
in~\cite{MUO-11-001}. 

Background studies play a decisive role with respect to the performance of the existing
detector, as well as to the design of the upgraded Muon system for the High Luminosity running
of the LHC (HL-LHC).
In general, background radiation levels are an important consideration in the overall 
performance of the  Muon system, and are carefully monitored. 
Any background source, including low energy neutrons or photons,
electrons and positrons, punch-through hadrons,  
low momentum primary and secondary muons, and beam-induced background
(particles produced in the interaction of the beams with collimators, 
residual gas, and beam pipe components), 
could affect the muon trigger performance and pattern recognition 
of muon tracks. 
In particular, spurious hits due to noise or to radiation background could promote
low transverse-momentum muons to higher momentum.
In addition, excessive radiation levels can cause aging of the detectors. 
Furthermore, the expected radiation levels during the HL-LHC
can contribute to drive the choice of the most suitable technology for the detector
upgrade.  Reliable estimates of the expected hit rates are therefore essential. 


\section{The CMS RPCs at the LHC}
\label{section:intro}

CMS is one of the six experiments 
currently operating at the LHC.
The central feature of the general-purpose
CMS detector 
is a superconducting solenoid,
of 6 m internal diameter, providing a field of 3.8 T. 
%
%
Muons with pseudorapidity in the range $| \eta | < 2.4 $ are measured 
in gas-ionization detectors embedded in the steel return yoke,
with detection planes made of three technologies: 
Drift Tube (DT) chambers, Cathode Strip Chambers (CSC)
and Resistive Plate Chambers (RPC). 

\begin{figure}[htp]
\centering
\includegraphics[width=.45\textwidth]{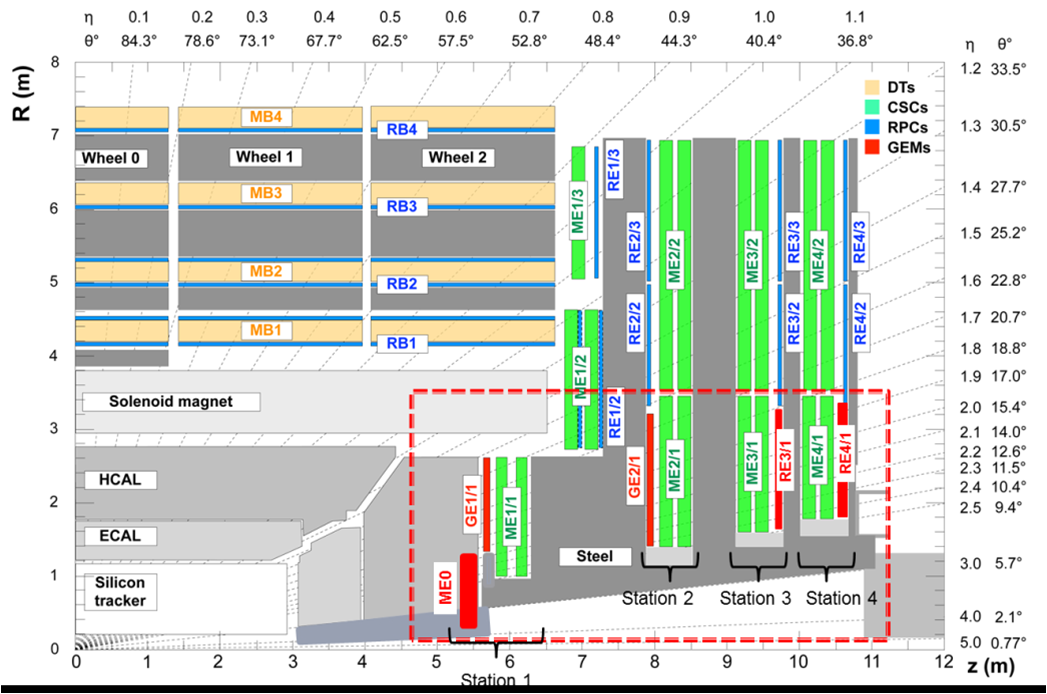}
\includegraphics[width=.35\textwidth]{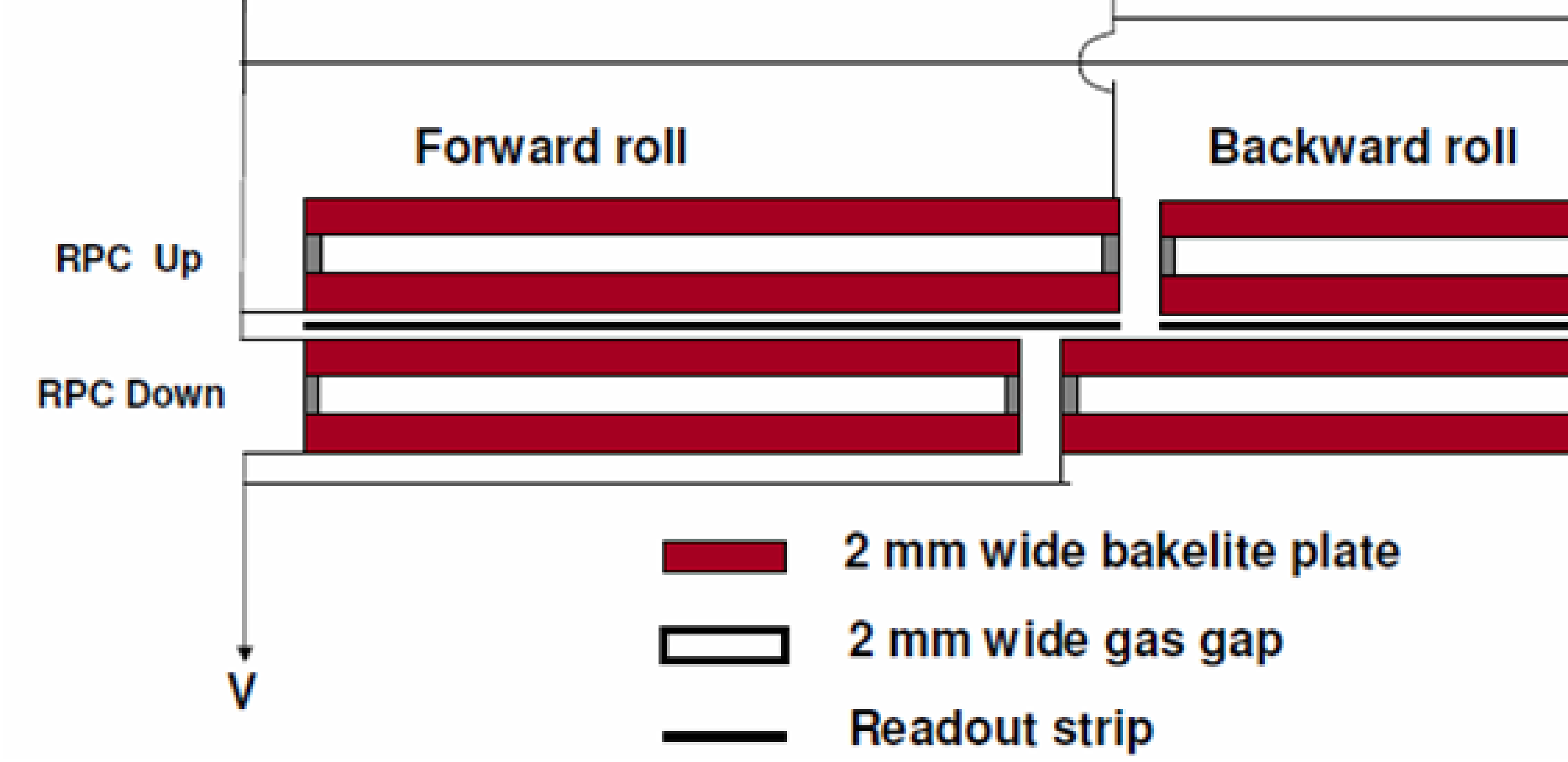}
\caption{Left: 
Schematic view, in the R-z plane, of one quadrant of the CMS detector,
with the axis parallel to the beam (z) running horizontally and the radius (R) 
increasing upward. The interaction region is at the lower left corner.
The position of the present RPC chambers is shown in blue. The RPCs
are both in the barrel (``RB'' chambers) and in the endcaps (``RE'') of CMS. 
The DT chambers are 
labeled ``MB''(``muon barrel'') and the CSC chambers are 
labeled ``ME'' (``muon endcap'').
The steel disks are displayed as dark gray areas.
Also shown in red are the chambers of the proposed upgrade scenario
of the CMS Muon system, including Gas Electron Multiplier detectors
(labeled ``ME0'' and ``GE'') and RPCs (``RE3/1'' and ``RE4/1''). 
Right: Schematic layout of a double-gap barrel chamber composed by two sub-units, called rolls.  
The readout strip plane is also shown. 
}
\label{fig:detector}
\end{figure}

The RPC~\cite{cms_rpc1, cms_rpc2} detectors are implemented in CMS 
as a dedicated trigger system, 
in both the barrel and the endcap regions. 
Figure~\ref{fig:detector} shows a schematic view of 
one quarter of the CMS detector
in the R-z plane (Fig.~\ref{fig:detector}, left)
and 
the layout of a double-gap barrel chamber 
(Fig.~\ref{fig:detector}, right).
Two gas gaps, of 2 mm each, are formed by two parallel phenolic-melaminic laminate (bakelite) 
electrodes, with
one single plane of copper readout strips  in-between. The two gaps feature
2 mm thickness and have a bulk resistivity of the order of $10^{10} \Omega$~cm.
High Voltage (HV) is applied to the outer graphite-coated surface of the bakelite plates.
The chambers are equipped with front-end boards, each one connected to 16 strips.

The barrel RPC system consists of five wheels,
installed at $| \eta |< $ 0.8 and $| \mathrm{z} |< 7 $~m,
subdivided into 12 azimuthal sectors, each one equipped 
with six radial layers of RPCs. The radial layers are referred to in the 
following as RB1, RB2, RB3 and RB4 
(two RB1 and two RB2 layers exist, for a total of six
radial layers). 
Six endcap disks, three on the positive and
three on the negative endcap side, are divided into 36 azimuthal sectors,
with two radial rings (labeled 
R2 and R3) in each one. 
They assure a full coverage 
up to $| \eta |< $ 1.6, as shown in Fig.~\ref{fig:detector}.
The fourth endcap disks, one on each side, are being installed during 
the 2014 shutdown and
were not present during the 2010-2012 data taking period.
The innermost rings (R1) will
be installed during a future shutdown.

In total, 480 barrel chambers and
432 endcap chambers were installed in the 2010-2012 data taking period,
adding up to 68136 barrel strips and
41472 endcap strips, respectively, and covering a total surface of about 3000 m$^2$. 
The CMS RPCs work in saturated avalanche mode and use a three-component,
non-flammable gas mixture composed of 95.2\% C$_2$H$_2$F$_4$ (R134a), 
4.5\% iC$_4$H$_{10}$ (isobutane) and 0.3\% SF$_6$.
After mixing, water vapor is added in order to maintain
the relative humidity at constant values of 40\%-50\% and to allow for 
constant bakelite resistivity.
%
The strip 
pitch dimensions are between 2.28 and 4.10~cm in the barrel and between 
1.95 and 3.63~cm in the endcap.

The RPCs have performed very well, with more than 98\% of the channels operational.
They were responsible for the loss of less than 1\% of CMS running time. The fraction 
of recorded pp collision data certified good by the RPCs was above 99\%.

In 2011~(2012), CMS has recorded 5.2~(21.8)~\fbi of data out
of 5.7~(23.3)~\fbi delivered by the LHC, for an efficiency of 91~(about 94)~\%. 
An average of 98\% of the CMS subdetector channels were operational and in the
readout. Roughly 93\%
of the recorded data have been certified as ``golden'' for all physics analyses.
Only certified runs have been used for the background studies.


%

\section{Background measurement technique}
\label{section:technique}

The RPC hit rate is measured at the strip level during LHC pp collision runs.
The  strip rate is calculated by using the incremental counts of the RPC trigger 
Link Boards. The incremental counts are taken during typical time 
intervals of the order of 100~s.
The resulting rates are then averaged over
the total runtime and normalized to the strip area. The instantaneous luminosity
is averaged over the same runtime. Normalized rates in Hz/cm$^{2}$ are shown
in this paper.

It is worth noting that no trigger selection is applied at this stage, resulting
in an inclusive measurement of the radiation background rates.
The offset level due to the intrinsic noise is estimated (if present) 
for each run and each 
chamber separately through a linear extrapolation 
to the value corresponding to zero instantaneous luminosity, 
which is subtracted from the chamber rate.
The average noise rate measured during 2010-2012 cosmic runs is of the order 
of 0.1~Hz/cm$^{2}$, as shown in Fig.~\ref{fig:noise} (left), {\it i.e.} much lower
than the average background rates shown in Section~{\ref{section:results}.

\begin{figure}[htbp]
\centering
\includegraphics[width=.39\textwidth]{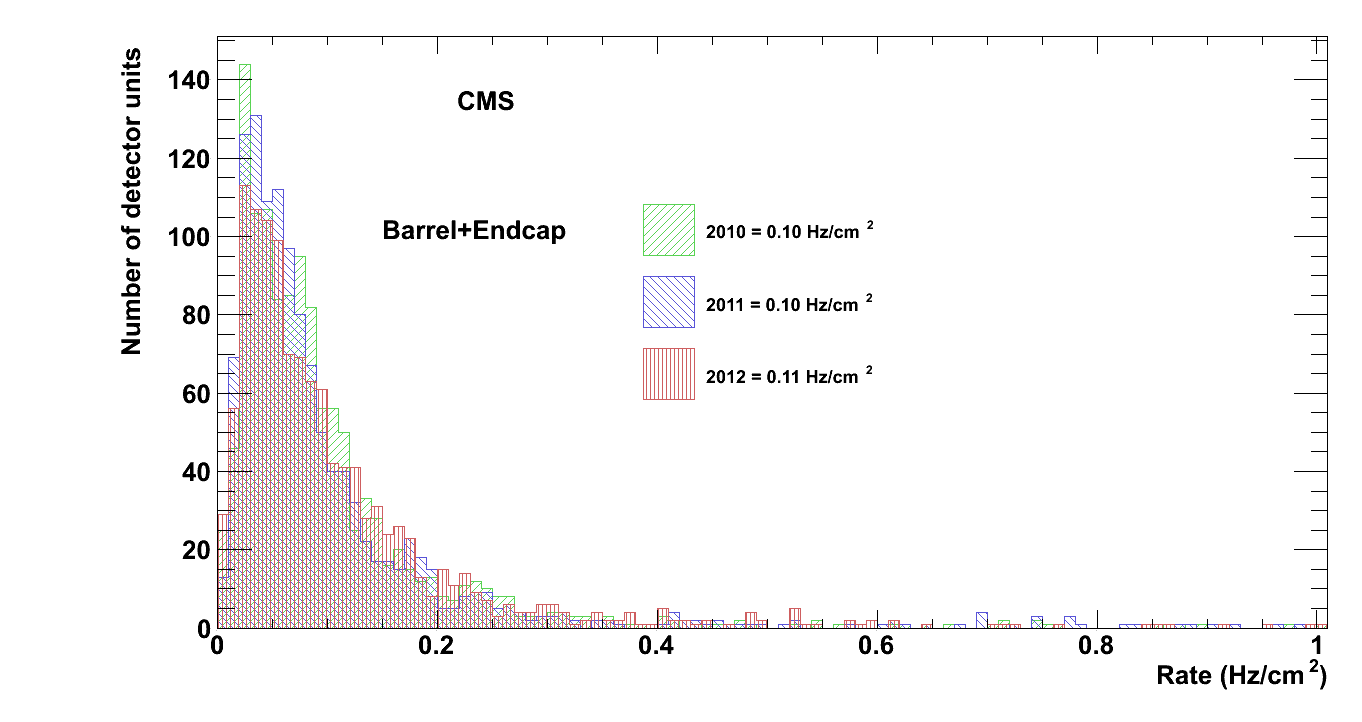}
\includegraphics[width=.39\textwidth]{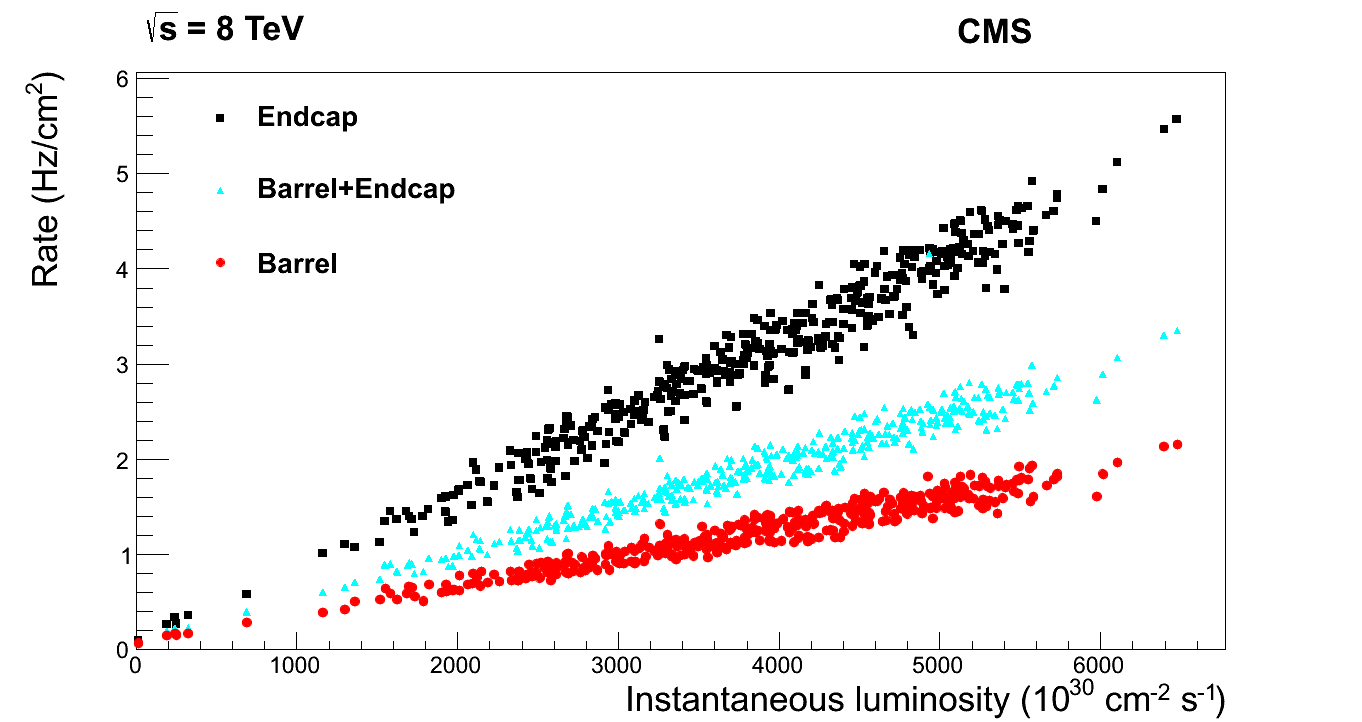}
\caption{
Left: average noise rate measured in the whole RPC detector, in 
2010 (light green), 2011 (blue), and 2012 (dark red) 
in absence of pp collisions, during cosmic runs. The average noise
level is approximately 0.1~Hz/cm$^{2}$, constant in time. 
Right: average background rate as a function of the
instantaneous luminosity, measured in 2012 in the RPC barrel (red)
endcap (black) and in the total system (barrel plus endcap, 
shown in light blue). 
}
\label{fig:noise}
\end{figure}

\section{Background measurement results}
\label{section:results}

\subsection{Linear dependence on instantaneous luminosity}

A linear relationship has been observed between the measured RPC hit rate
and the LHC instantaneous luminosity, for several orders of magnitude of
the instantaneous luminosity, from $10^{29}$~cm$^{-2}$~s$^{-1}$ in 2010,
to almost $10^{34} $ cm$^{-2}$~s$^{-1}$ in 2012.  
The linear relationship 
holds for every single chamber, with correlation coefficient greater than 95\%
in more than 80\% of the chambers, and also for larger detector parts (layers, rings, disks, wheels, etc.), 
as displayed in the following
Figures showing the measured rate as a function of the luminosity. 

No deviation from linearity has been observed in the 2010-2012 range of the LHC
luminosity,
therefore excluding, in this range, phenomena which could give rise to power-law
dependence relationships with exponent different from 1.
Extrapolations from the present measurements to larger luminosity values have thus
been performed linearly and are reported in Section~\ref{section:extrap}.
%
Figure~\ref{fig:noise} (right) shows the average background rates measured in the RPC
detector as a function of the instantaneous luminosity.
Larger average rates are measured in the endcap, which spans a 
higher $| \eta |$ region.

\subsection{Background as a function of $z$ and plus-minus asymmetry}
\label{subsection:z}

In the central $\eta $ region the rates increase as the chambers are further
along z from the interaction point (IP),
with the largest rates observed in the 
outermost
wheels $W+2$ and $W-2$~(Fig.~\ref{fig:barrel}), 
which are the ones most exposed 
to the slow neutron gas permeating the cavern. 
The asymmetry
between the two external wheels, with larger rate measured on the 
positive side~($W+2$) with respect to the negative side~($W-2$),
might be due to lower density of the neutron gas on the negative
side because of 
the presence of the CMS main shaft (through which 
the detector was lowered 
from the surface into the cavern).

The same pattern is observed at high $| \eta |$~(Fig.~\ref{fig:endcap}),
with higher rate as $|\eta|$
increases, {\it i.e. } in the chambers located further away from the IP.
Both the inner (R2) and the outer (R3) rings of the six endcap disks 
show an increasing background as $z$ increases.

It is interesting to note the asymmetry between the rings
R2 of the negative (labeled ``D-2 R2'') and the positive 
(labeled ``D+2 R2'') second disk of the endcap, 
due to missing shielding tiles on the 
negative endcap yoke 
supporting the disk D-2.

\begin{figure}[hbtp]
\centering
\includegraphics[width=.39\textwidth]{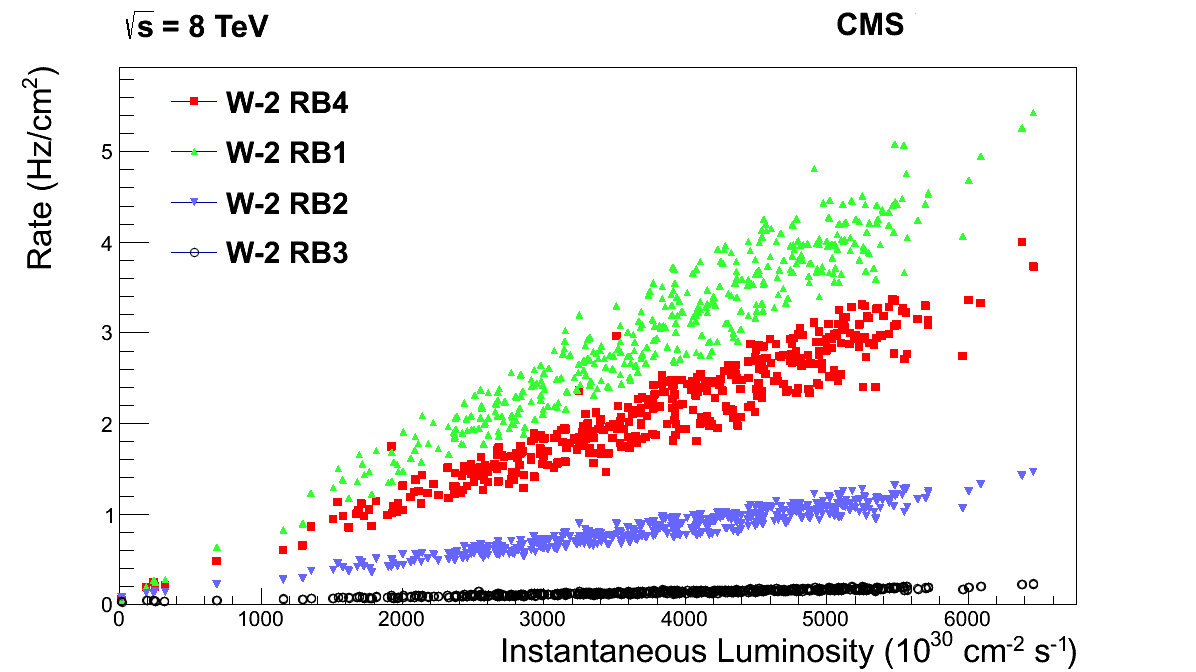}
\includegraphics[width=.39\textwidth]{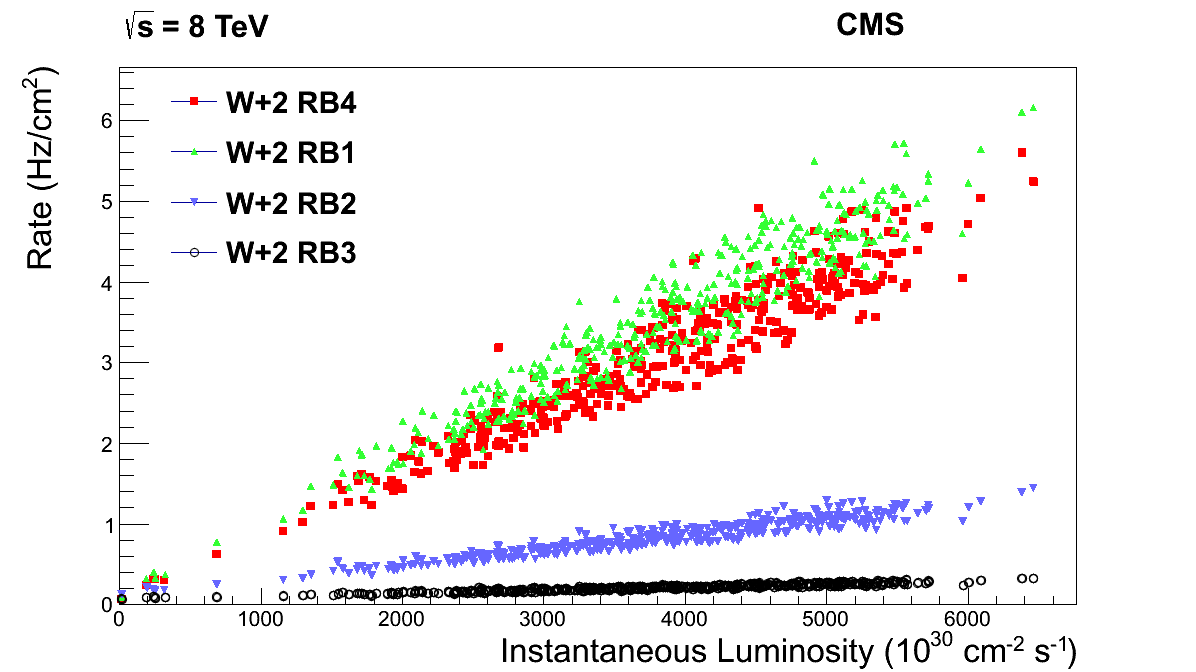}
\caption{
Background rate 
in the radial layers of W-2 (left)
and W+2 (right). Larger rates are measured in the innermost (RB1)
and outermost (RB4) layers, as discussed in the text. The plus-minus
asymmetry of the RB4 layers of the external wheels can also be 
appreciated from the comparison between the two
plots, showing higher background levels on the plus side.
}
\label{fig:barrel}
\end{figure}

\begin{figure}[hbtp]
\centering
\includegraphics[width=.39\textwidth]{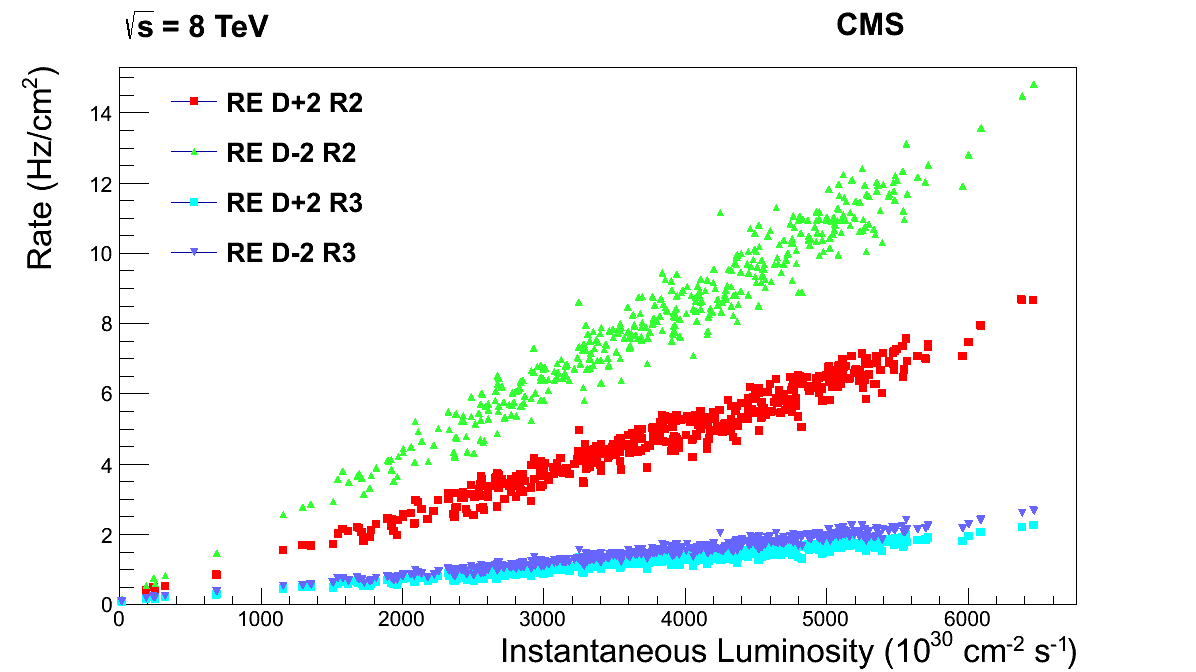}
\includegraphics[width=.39\textwidth]{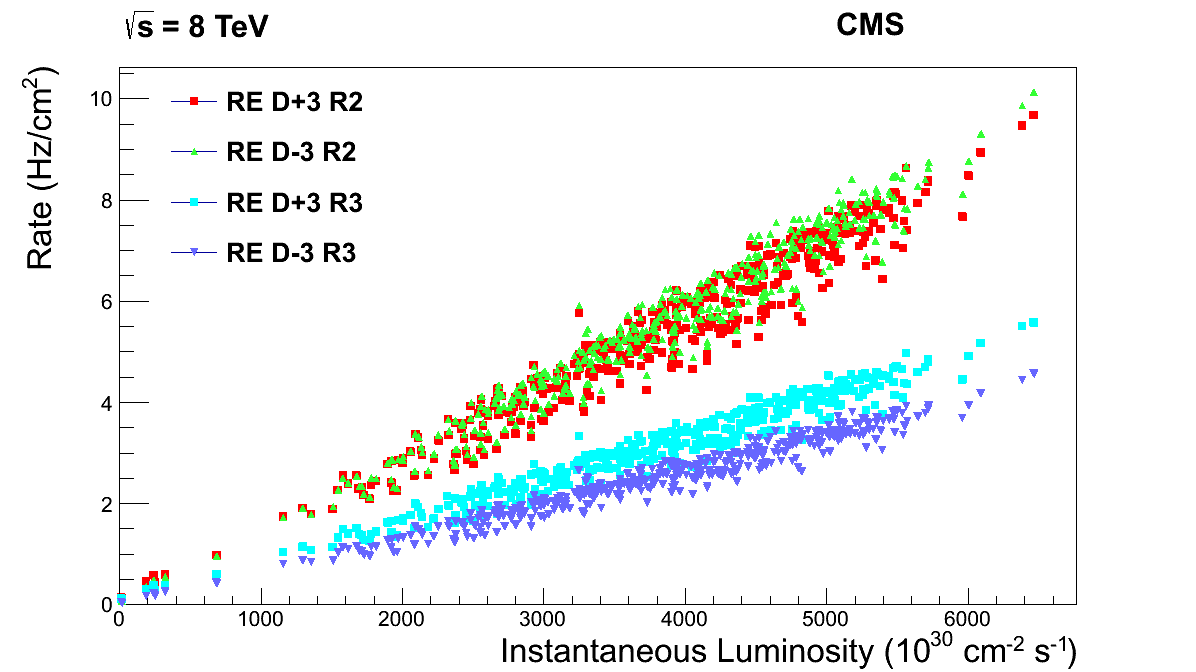}
\caption{
Background rate 
in the disks D-2 and D+2 (left)
and D-3 and D+3 (right). For each disk, the two radial rings R2 and
R3 are shown.   
}
\label{fig:endcap}
\end{figure}


\subsection{Background as a function of $R$}

The background behavior as a function of $R$ is different in the barrel
and in the endcap, somehow reflecting the different types of background 
sources near the beamline and on the outside of CMS.

In the barrel~(Fig.~\ref{fig:barrel}), the largest background 
is observed both in the inner layers (RB1),
and in the outer layers (RB4). The innermost chambers are exposed to
particle leakage from the hadron calorimeter, and from the gap between
the barrel and the endcap of the calorimeters, while
the barrel outermost chambers are mostly exposed 
to the surrounding radiation background from slow neutrons.
Chambers located in the central layers RB2 and RB3
are protected by the steel of the barrel wheels and therefore detect 
much lower rates.

In the endcap~(Fig.~\ref{fig:endcap}), the highest rates are detected
in the inner rings (rings R2 of all endcap disks). The radial rings R2
are the closest to the beamline. The background rates 
decrease along $R$ as the distance from the beamline increases.
The outer (R3) chambers show a smaller increase in rate with luminosity
relative to the inner (R2) chambers.

\subsection{Background as a function of $\phi$ and azimuthal asymmetry}

A strong azimuthal dependence of the background
has been observed in the outer barrel
layers (RB4) of all barrel wheels~(Fig.~\ref{fig:barrelphi}), while
in the inner layers the background distribution is symmetrical.
In RB4 the rate ratio between the top and the bottom sectors is
approximately a factor 20.
The top-bottom asymmetry is explained by the non-symmetric features 
of the -otherwise symmetric- CMS detector: the concrete and steel flooring
and the wheel and disk supports.

No evidence of $\phi$-asymmetry is observed in the 
endcap~(Fig.~\ref{fig:barrelphi}), most likely because 
the endcap chambers are located in between steel disks, while the
barrel outer top chambers (RB4, top $\phi$ sectors from 3 to 5) 
are more exposed to the surrounding background.

\begin{figure}[hbtp]
\includegraphics[width=.28\textwidth]{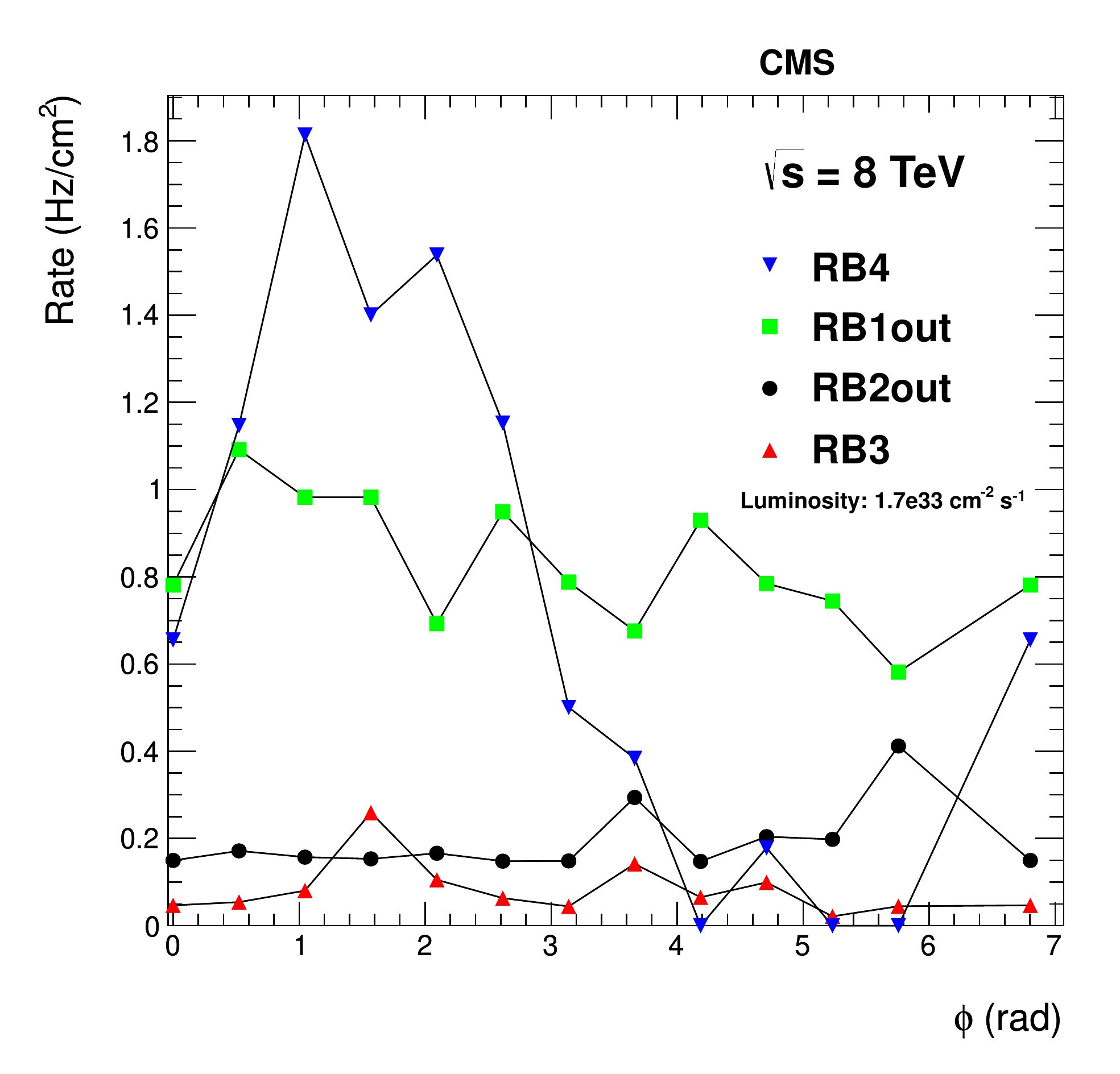}
\includegraphics[width=.28\textwidth]{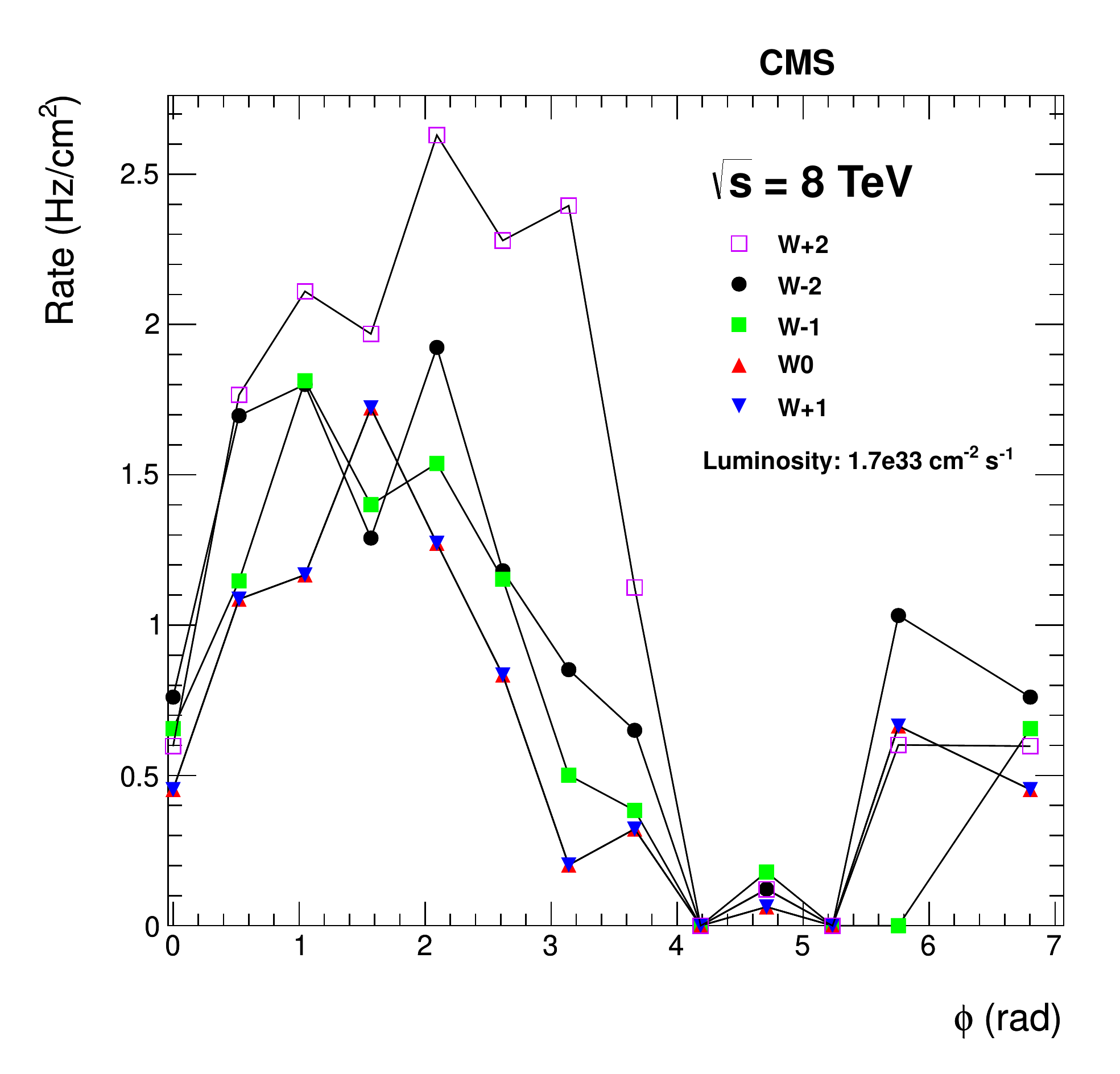}
\includegraphics[width=.45\textwidth]{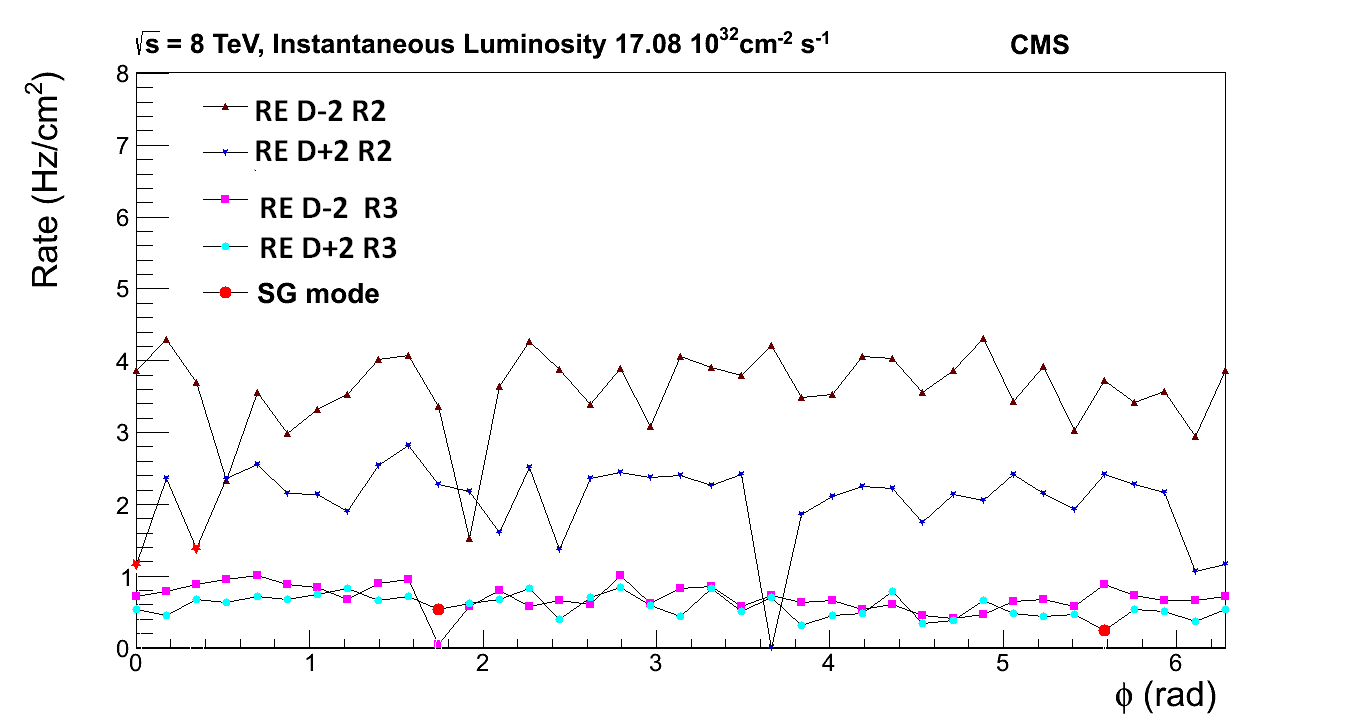}
\caption{
Left: Background rates
of RPC chambers located
in inner and outer layers of the barrel  wheel W-1, 
as a function of the chamber azimuthal
position, for a luminosity of  
$1.7 \cdot 10^{33}$ cm$^{-2}$ s$^{-1}$ in 2012. A dependence on
the chamber azimuthal position ($\phi$ asymmetry) is shown for the external
layer RB4.
Middle: Background rates of the RB4 layers in all barrel wheels.
The larger rate in the W+2 wheel is discussed in the text.
Right: 
Background rates
of RPC chambers located
in the rings R2 and R3 of the endcap disks D+2 and D-2, 
as a function of the chamber azimuthal
position, for a luminosity of  
$1.7 \cdot 10^{33}$ cm$^{-2}$ s$^{-1}$ in 2012. Note the higher rate
in D-2/R2, mentioned in the text.
}
\label{fig:barrelphi}
\end{figure}


\subsection{Background distribution: summary maps}

The background measurement results described in details in the previous
sections are also visually summarized in the two-dimensional plots
shown in Figs.~\ref{fig:map1} and~\ref{fig:map2}. 
In the barrel, larger rates are measured in the top (3,4,5)
sectors of RB4, and in the internal radial layer RB1, independently 
of the sector. In the endcap, the rates are higher close to the beamline.
The maps show no evidence of $\phi$ asymmetry 
in the endcap.

\begin{figure}[hbp]
\centering
\includegraphics[width=.39\textwidth]{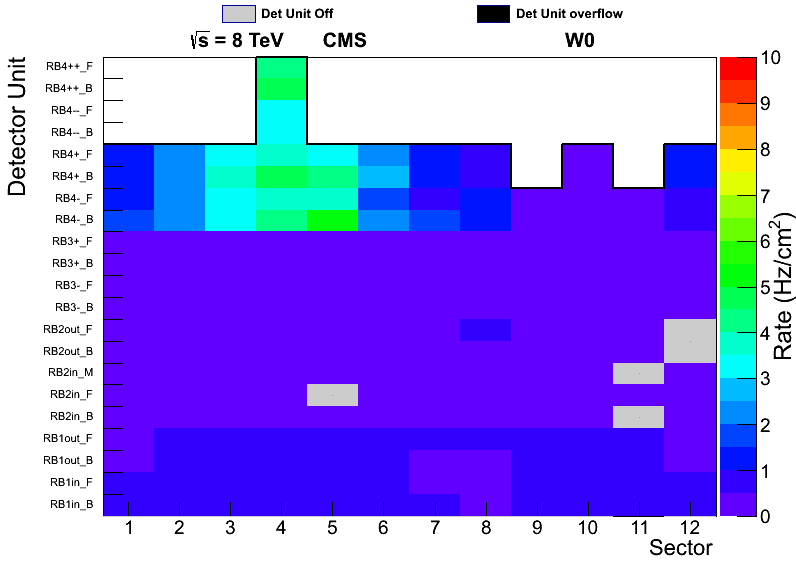}
\includegraphics[width=.39\textwidth]{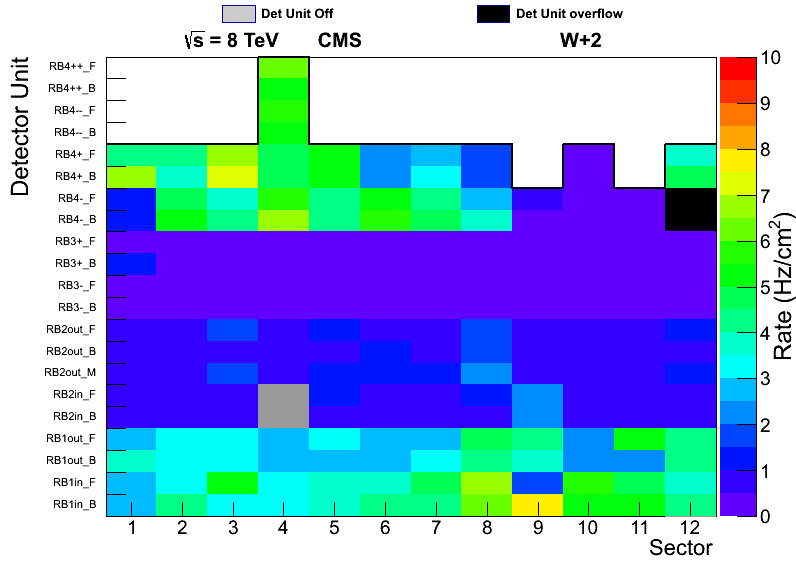}
\caption{
Two-dimensional maps of the background distribution in the central
barrel Wheel 0 (left), and in the external positive Wheel+2, where the
highest rate is detected. Each box represents a detector
unit in the $R-\phi$ plane. The chambers are ordered from RB1 to RB4,
for increasing radial distance from the beamline, and along the 
azimuthal trigger sectors. 
}
\label{fig:map1}
\end{figure}

\begin{figure}[hbp]
\centering
\includegraphics[width=.39\textwidth]{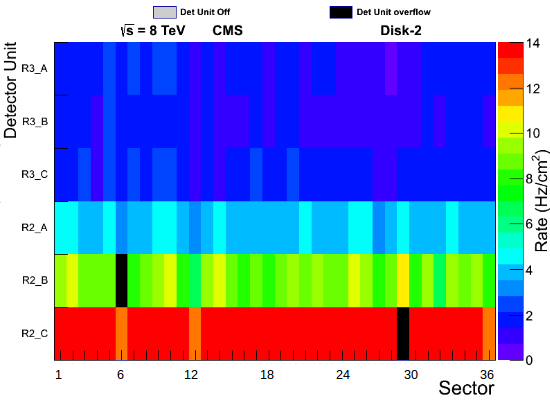}
\includegraphics[width=.39\textwidth]{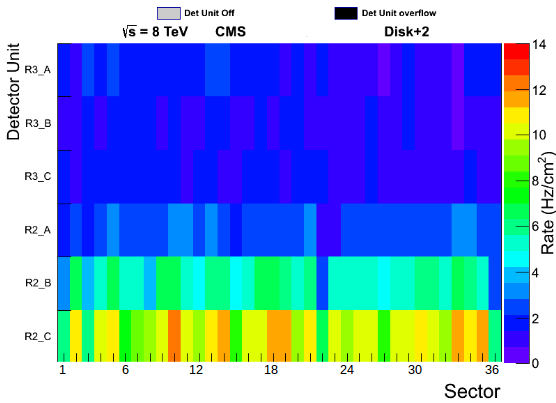}
\caption{
Two-dimensional maps of the background distribution in the 
endcap Disk-2, where the
highest rate is detected (as explained in the text),
and in the Disk+2.
Each box represents a detector
unit in the $R-\phi$ plane. The chambers are ordered radially in $\eta$
partitions with increasing radial distance from the beamline (with
partition C of ring R2 being the closest to the beamline),
and along the 
azimuthal trigger sectors. 
}
\label{fig:map2}
\end{figure}

\subsection{Comparison with the CMS CSCs and DTs}

Similar features have been measured also by the CSCs and DTs,
including the linear dependence on the LHC instantaneous luminosity
and the typical patterns of the radiation background distributions.
A comparison with 2010 data among the three detectors of the CMS 
Muon system was performed in~\cite{MUO-11-001}.

\section{Extrapolation to higher luminosities}
\label{section:extrap}

Based on the linear relationship between measured rates and 
instantaneous luminosity, the RPC rates can be linearly extrapolated
to higher luminosity values. 

The extrapolation to higher beam energy
(from 7 or 8~TeV of the present pp data to 14~TeV) can be performed
by rescaling by an approximate total factor of 1.5, due to  
larger cross section and particle multiplicity: the pp 
inelastic cross section increases from 73.5~mb~\cite{totem} at 7~TeV
to 80~mb~\cite{epos} at 14~TeV; a factor $\sqrt{2}$ accounts for
the increased particle
multiplicity, scaling as square root of the energy.

At a peak luminosity of $10^{34}$~
(or $5 \cdot 10^{34}$)~cm$^{-2}$ s$^{-1}$, and at the 
center-of-mass energy of 14~TeV, the expected average rates 
are of the order of 4 (20)~Hz/cm$^{2}$ in the barrel and 
12 (60)~Hz/cm$^{2}$ in the endcap. The maximum rates in the most
exposed regions of the RPCs are within a factor 5 with respect
to the quoted average rates, and are therefore well within the RPC rate
capability, which is of the order of 1~KHz/cm$^{2}$~\cite{capability}

\section{Comparison with simulation}

Comparisons with FLUKA~\cite{fluka} simulations have shown that
the simulation describes well the typical patterns of the
radiation background inside the RPCs. 
Figure~\ref{fig:fluka} (left) shows, as an example, the neutron 
flux inside the Muon system. As measured in the data, the flux 
is expected to be high in the RB1 and RB4 barrel layers, and in
the endcap at high $|\eta|$. 
The reference luminosity of $10^{34}$ cm$^{-2}$ s$^{-1}$ is assumed.

Flux and rate estimates from simulation can be used in the 
high $|\eta|$ regions -not covered by the present Muon system-
where additional RPC chambers will be installed in a future
shutdown.
In order to provide estimates of the expected rates, the FLUKA
flux values have been rescaled by the following average sensitivity 
values~\cite{sensitivity} from previous simulations and test beam
measurements: 0.001 for neutrons, 0.01 for photons, and 1 for charged
particles. Neutrons, photons, and charged particles account for
almost 100\% of the particles produced in primary pp interactions. 
In total, a rate of a few tens of Hz/cm$^{2}$ is expected at 
low~R and high $|\eta|$ values, {\it i.e.} in the region closest
to the beamline (Fig.~\ref{fig:fluka}, right).

\begin{figure}[htbp]
\centering
\includegraphics[width=.47\textwidth]{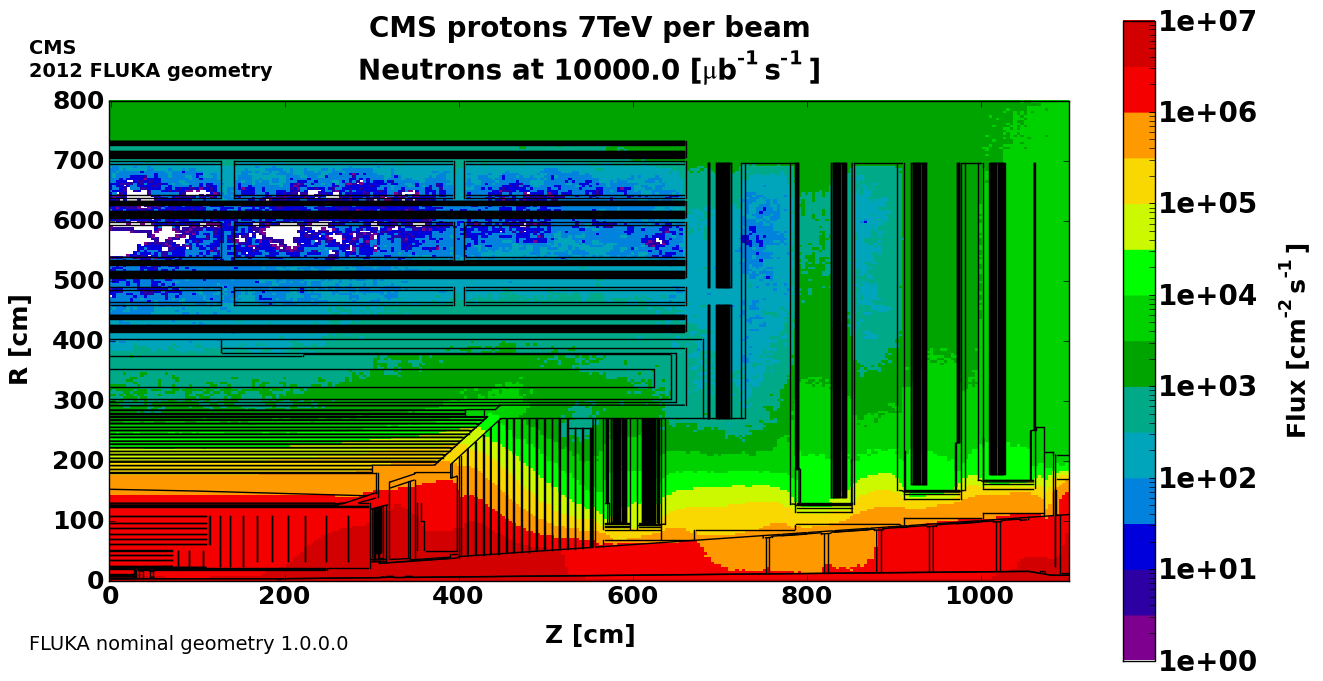}
\includegraphics[width=.37\textwidth]{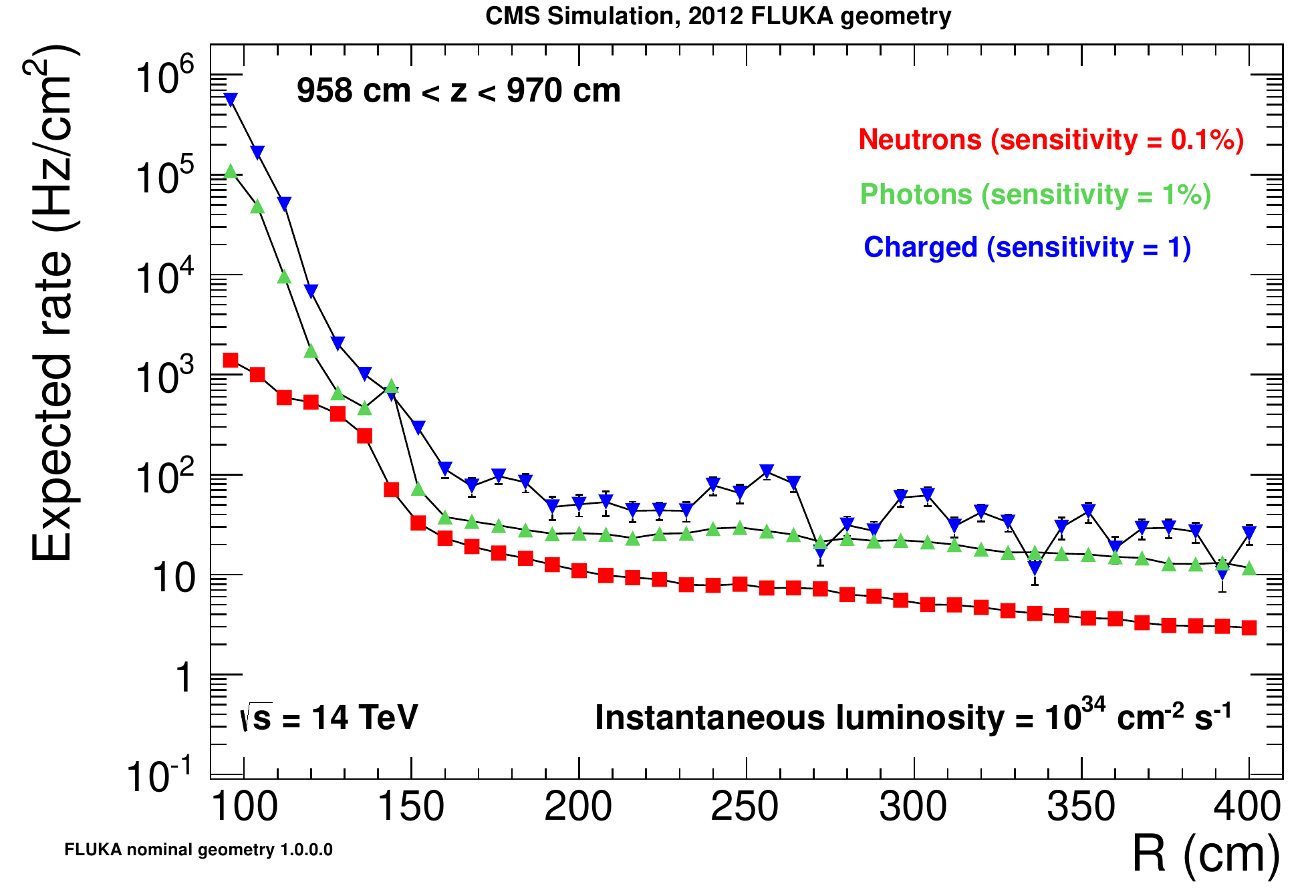}
\caption{
Left: Expected neutron flux, in Hz/cm$^{2}$, in the Muon system.
An R-z view of one quarter of the CMS detector is shown.
Right: Expected rate, from neutrons, photons, and charged particles,
as a function of the radial distance from the beamline, for 
a z distance between 790 and 802~cm from the IP. The future RE3/1 
ring will cover the radial range between 160 and 330~cm.  
}
\label{fig:fluka}
\end{figure}

\section{Conclusions}

Detailed studies of the radiation background have been
performed, which have lead to an accurate understanding 
of the background distribution inside the CMS Muon system.
Comparisons with simulations have shown reasonable agreement
between the measurements and the expected rates. 
Extrapolations to high-luminosity scenarios predict background
levels within the RPC rate capability.

%

\acknowledgments

Acknowledgments of support of all CMS are given in Ref.~\cite{MUO-11-001}.
The authors would like to thank the RPC2014 organizers and 
the colleagues of the CMS Muon system.

\end{document}